\documentclass[twocolumn]{article}

\usepackage{PRIMEarxiv}

\usepackage{times}
\usepackage{subcaption}
\usepackage{booktabs}
\usepackage{amsmath,amssymb,amsfonts}
\usepackage{mathtools}
\usepackage{siunitx}
\usepackage[numbers,sort&compress]{natbib}
\usepackage{hyperref}
\usepackage[nameinlink,capitalize]{cleveref}
\usepackage{xcolor}
\usepackage{enumitem}
\usepackage{orcidlink}
\usepackage[utf8]{inputenc}
\usepackage[T1]{fontenc}
\usepackage{url}
\usepackage{nicefrac}
\usepackage{microtype}
\usepackage{fancyhdr}
\usepackage{graphicx}
\usepackage{listings}
\graphicspath{{media/}}
\usepackage{lipsum}
\usepackage{xfrac}
\usepackage{makecell}
\usepackage{multirow}
\usepackage{comment}
\usepackage{xcolor}
\usepackage{ulem}
\usepackage{float}
\graphicspath{{figures/}}

\newcommand{\Cp}{C_p}
\newcommand{\Froude}{\mathrm{Fn}}
\newcommand{\Lpp}{L_{\mathrm{PP}}}
\newcommand{\Bwl}{B_{\mathrm{WL}}}
\newcommand{\Km}{K_M}

\title{ShipNet: A Geometric Deep Learning Surrogate for Real-Time Ship Hydrodynamics}

\author{
 Kirsten Odendaal \\
  Maritime Research Insitute Netherlands\\
  Wageningen, Netherlands \\
  \texttt{k.odendaal@marin.nl} \\
   \And
 George Drakoulas \\
  Damen Research\\
   Gorinchem, Netherlands\\
  \texttt{george.drakoulas@damen.com} \\
}

\date{} 

\begin{document}

\twocolumn[{
  \begin{@twocolumnfalse}
  
\maketitle
\begin{abstract}
\vspace{5pt}
Accurate prediction of hydrodynamic performance is central to ship design, yet high-fidelity computational fluid dynamics remains prohibitively expensive for large-scale parametric exploration. This motivates the development of data-driven surrogate models that provide rapid approximations to hydrodynamic predictions at substantially reduced cost. We present \textit{ShipNet}, a geometric deep-learning surrogate that predicts both hull-surface pressure distributions and far-field free-surface wave patterns directly from hull geometry and speed. The network employs a regularized dynamic graph convolutional backbone on hull point clouds, with a multi-head decoder for simultaneous near-body pressure and free-surface elevation outputs. Training data consist of 420 inviscid free-surface simulations generated using a potential-flow panel method for two parent yacht hulls, each parameterized into 70 variants and evaluated at three speeds. \textit{ShipNet} predicts per-point pressure coefficient and two-dimensional wave elevation map using a composite loss that combines point-wise regression and image-structure terms. On a geometry-held-out test set, \textit{ShipNet} achieves $R^2\approx0.98$ for hull pressure and $R^2\approx0.91$ for wave fields.
Inference requires $\approx0.15 s$ per case, yielding over a $550\text{x}$ speedup relative to the potential-flow solver on conventional hardware. Limitations include the restricted geometry and speed ranges and the inviscid training data, while future work will extend the model to high-fidelity viscous simulations with physics-informed regularization.
\end{abstract}
  \end{@twocolumnfalse}
  \vspace{2em}
}]

\section{Introduction}
The maritime industry is undergoing a shift toward more sustainable and efficient operation, driven by tightening environmental regulation and economic pressure. A central lever is hull-form design, since the hull strongly influences hydrodynamic performance and therefore fuel consumption 
\cite{TadrosHullReview}. In modern practice, computational fluid dynamics (CFD), particularly Reynolds-Averaged Navier-Stokes (RANS) simulations, is widely used to evaluate candidate designs by resolving hull pressure and the associated free-surface wave pattern \cite{larsson2010ship}. These quantities are directly linked to resistance components and are therefore valuable targets for early-stage screening.

Despite its accuracy, CFD is difficult to use as an interactive design tool. High-fidelity simulations can require substantial computational resources and long turnaround times, especially when free-surface effects, trim, and sinkage are included \cite{vanDerKolk2020,HuangFullScaleGuideline}. In addition, reliable CFD results require expertise in geometry preparation, meshing, solver settings, and model selection. These practical constraints reduce the number of designs that can be evaluated and slow down the design spiral.

Data-driven surrogate models aim to improve this speed and accuracy trade-off by learning a direct mapping from design inputs to flow outputs \cite{drakoulas2024explainable}. Classical reduced-order modeling has been successful for parametric variation in boundary conditions and material properties, but typically assumes fixed spatial discretizations and can struggle to accommodate geometric changes \cite{bennerROM, drakoulas2023fastsvd}. Recent scientific machine learning methods have expanded surrogate modeling to field prediction, including encoder-decoder approaches, physics-informed training, and operator learning \cite{drakoulas2024physics,RaissiPINN,LiFNO,LuDeepONet}. However, many approaches still require structured grids or implicit geometry encodings that are not ideal when the geometry itself is the dominant source of variability.

Geometric deep learning provides an attractive alternative because it operates directly on unstructured representations such as point clouds and meshes, allowing the model to learn features that align with local geometric structure \cite{BronsteinGDL}. In external aerodynamics, regression-based Dynamic Graph Convolutional Neural Networks (regDGCNN), as demonstrated in DrivAerNet++, have shown that dynamic graph construction and EdgeConv features can learn meaningful geometry-conditioned representations from point clouds \cite{WangDGCNN, Elrefaie_2025, elrefaie_2025dataset}. This motivates adapting similar architectures to ship hydrodynamics, where geometry also governs pressure trends and wave generation.

This work adapts regDGCNN from single-phase automotive aerodynamics to two-phase ship hydrodynamics. The transition is non-trivial because ship flows are shaped by the coupled air-water interface and by the vessel response (trim and sinkage). Our core hypothesis is that a geometry-aware backbone that learns local relationships between hull shape and pressure can be extended to a shared latent representation that supports joint prediction of both hull-surface pressure and the free-surface wave elevation field. This multi-task formulation targets both near-body and far-field behavior, and it aligns with early-stage design needs where fast ranking and qualitative insight are often more important than exact viscous detail.

\textbf{Contributions.} The primary contributions of this paper are:
\begin{enumerate}[topsep=0pt, itemsep=0.0pt, leftmargin=*]
    \item An adaptation of a dynamic-graph CNN (regDGCNN) to predict ship hydrodynamic fields from parametric hull geometries and operating conditions.
    \item A multi-head formulation that jointly predicts hull-surface pressure (a 3D manifold signal on the point cloud) and free-surface wave elevation (a 2D field) from a shared latent representation.
    \item A validated surrogate achieving high accuracy ($R^2 \approx 0.98$ for $C_p$, $R^2 \approx 0.91$ for $\eta$) with $\sim1500\times$ speedup versus the RAPID potential-flow solver used for data generation, enabling quick early-stage design exploration.
\end{enumerate}

The remainder of this paper is structured as follows. Section~2 reviews related work in both traditional and data-driven hydrodynamics. Section~3 details the methodology, including the dataset generation, architectural design, and training procedure. Section~4 presents results and analysis. Section~5 concludes with limitations and future directions.

\section{Methodology}
\subsection{Data and design space}
Two parent generic yachts, one with a straight bow and one with a bulbous bow, were parametrically varied using a Latin hypercube design, yielding 70 variants per family (140 total). Principal particulars are in \cref{tab:parents}. Extreme parameter ranges are in \cref{tab:doe}.

\renewcommand{\arraystretch}{1.5}
\begin{table}[t]
\centering
\captionsetup{width=0.95\linewidth}
\footnotesize
\resizebox{\columnwidth}{!}{%
\sisetup{detect-weight=true,detect-inline-weight=math}
\begin{tabular}{l l l S[table-format=2.2] S[table-format=2.2]}
\toprule
\textbf{Description} & \textbf{Symbol} & \textbf{Unit} & \textbf{Straight} & \textbf{Bulbous} \\
\midrule
Length perpendiculars & $\Lpp$ & m & 99.36 & 99.36 \\
Breadth on waterline & $\Bwl$ & m & 15.50 & 15.50 \\
Mean draft & $T$ & m & 4.25 & 4.25 \\
Displacement volume & $\mathrm{DISP}_V$ & m$^3$ & 3535.88 & 3670.91 \\
Block coefficient & $C_B$ & -- & 0.54 & 0.56 \\
Midship section coef. & $C_M$ & -- & 0.82 & 0.82 \\
Prismatic coefficient & $C_P$ & -- & 0.66 & 0.68 \\
FP to buoyancy center & $\mathrm{FB}$ & \%$\Lpp$ & 53.77 & 51.82 \\
Transverse metacentre & $\Km$ & m & 8.63 & 8.39 \\
\bottomrule
\end{tabular}
}
\vspace{5pt}
\caption{Parent designs: main dimensions and hydrostatics.}
\label{tab:parents}
\vspace{-10pt}
\end{table}


\renewcommand{\arraystretch}{1.25}
\begin{table}[t]
\centering
\captionsetup{width=0.95\linewidth}
\footnotesize
\resizebox{\columnwidth}{!}{%
\begin{tabular}{@{}p{0.25\linewidth} p{0.75\linewidth}@{}}
\toprule
\textbf{Configuration} & \textbf{Variations} \\ \hline
\multirow{3}{*}{\text{Straight bow}} & 1. Stem extend: $x=99.3$ $\to$ $109.6$\,m (straight) \\
 & 2. Entrance angle: $0^\circ\to-7.5^\circ$ \\
 & 3. Fore shoulder: $0\to+0.5$\,m \\
\midrule
\multirow{5}{*}{\text{Bulbous bow}} & 1. Bulb length: $x=103.0\to 107.8$\,m \\ 
& 2. Bulb width: $y=1.20\to1.62$\,m \\ 
& 3. Bulb height: $z=3.50\to4.30$\,m \\
& 4. Entrance angle: $0^\circ\to-7.5^\circ$ \\ 
& 5. Fore shoulder: $0\to+0.5$\,m \\
\bottomrule
\end{tabular}
}
\vspace{5pt}
\caption{DOE extremes for bow-related parameters (normalized endpoints shown textually).}
\label{tab:doe}
\end{table}

\begin{figure}[t]
\centering
    \begin{subfigure}[b]{0.95\columnwidth}
        \centering
        \includegraphics[width=0.95\linewidth]{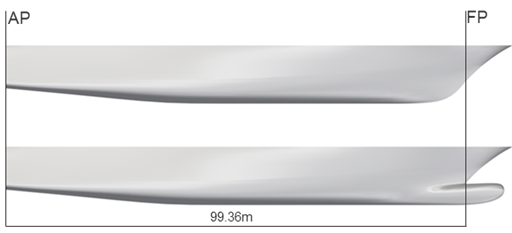}
        \caption{\label{fig:parenthulls} Geometry Profile}
    \end{subfigure}
    \quad
    \begin{subfigure}[b]{0.95\columnwidth}
        \centering
        \includegraphics[width=0.95\linewidth]{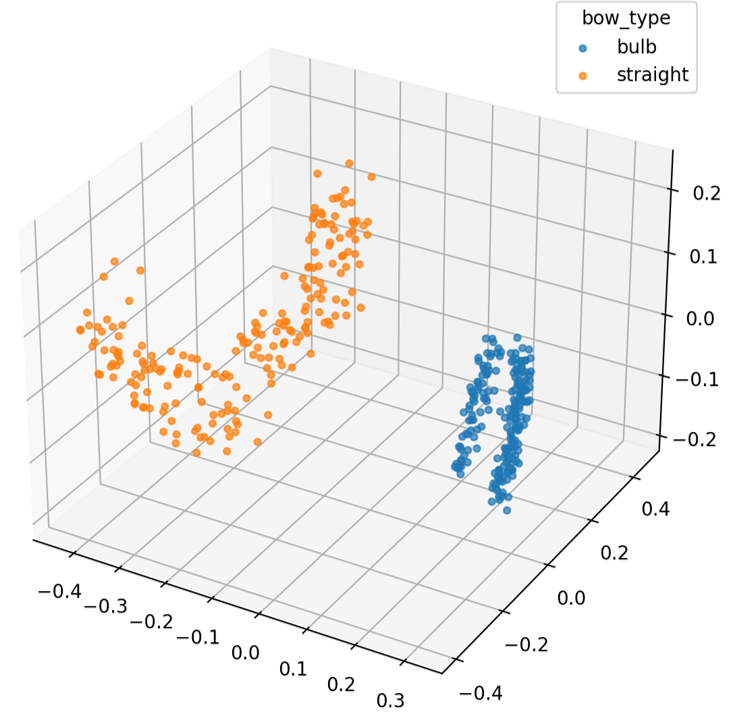}
        \caption{\label{fig:pcarep} Clustering Distributions}
    \end{subfigure}

    \caption{(a) Renderings of the two parent hulls, Yacht A1 and Yacht A2. (b) A PCA plot of the final 140 unique geometries, colored by parent class.}
    \label{fig:dataset_viz}
    \vspace{-5pt}
\end{figure}

The corresponding parent hull profiles and geometry variational clusters are shown in \Cref{fig:dataset_viz}. To visualize the coverage of the design space, we construct a coarse geometric descriptor for each hull by binning surface points into a fixed $16\times16\times16$ voxel grid over a shared spatial domain. The resulting occupancy histogram is normalized, flattened, and reduced using Principal Component Analysis (PCA). The first three components, explaining 88.9\% of the variance, are used solely for visualization and clustering analysis; this representation is not used as input to the learning model. 

\subsection{Simulation setup}
A robust and diverse dataset is crucial for training a generalizable surrogate model. Our dataset was generated using the potential flow panel method code RAPID \cite{RAPIDdoc}. This solver was selected as it provides an optimal trade-off between computational speed and physical fidelity for early-stage design exploration, allowing for the generation of a large dataset. While it is an inviscid method, its predictions of pressure distribution and wave-making characteristics are well-suited for ranking design candidates and have been shown in prior studies to correlate well with higher-fidelity RANS predictions for wave-making trends and relative design ranking \cite{Raven1996}. A consistent numerical setup was employed across all hull variants and speeds to ensure comparability. 

All 140 geometries were simulated at \SIlist{12;15;18}{kn} ($\Froude\in[0.197,0.322]$) using RAPID with consistent panelling and domain settings (\cref{tab:rapid}). For each run, we exported hull-surface $\Cp$ and a free-surface elevation map on a $256\times256$ grid. We split data into train/val/test as 70/15/15 with geometry-grouped splits to avoid leakage across speeds.

\renewcommand{\arraystretch}{1.25}
\begin{table}[t]
\centering
\captionsetup{width=0.95\linewidth}
\footnotesize
\resizebox{\columnwidth}{!}{%
\begin{tabular}{@{}p{0.6\linewidth} l @{}}
\toprule
\textbf{Parameter} & \textbf{Value} \\ 
\midrule
Hull panels (bulb/straight) & 6k / 2k (Quad) \\ 
Free-surface strips (half-width) & 34  \\ 
Free-surface panels (along hull) & 79  \\ 
Domain upstream / downstream & $0.50\Lpp$ / $0.807\Lpp$  \\ 
Domain half-width & $0.748\Lpp$ \\ 
Dynamic trim treatment & Enabled \\ 
\bottomrule
\end{tabular}
}
\vspace{5pt}
\caption{RAPID setup (common across cases).}
\label{tab:rapid}
\vspace{-15pt}
\end{table}

\begin{figure*}[t]
\centering
\includegraphics[width=0.95\linewidth]{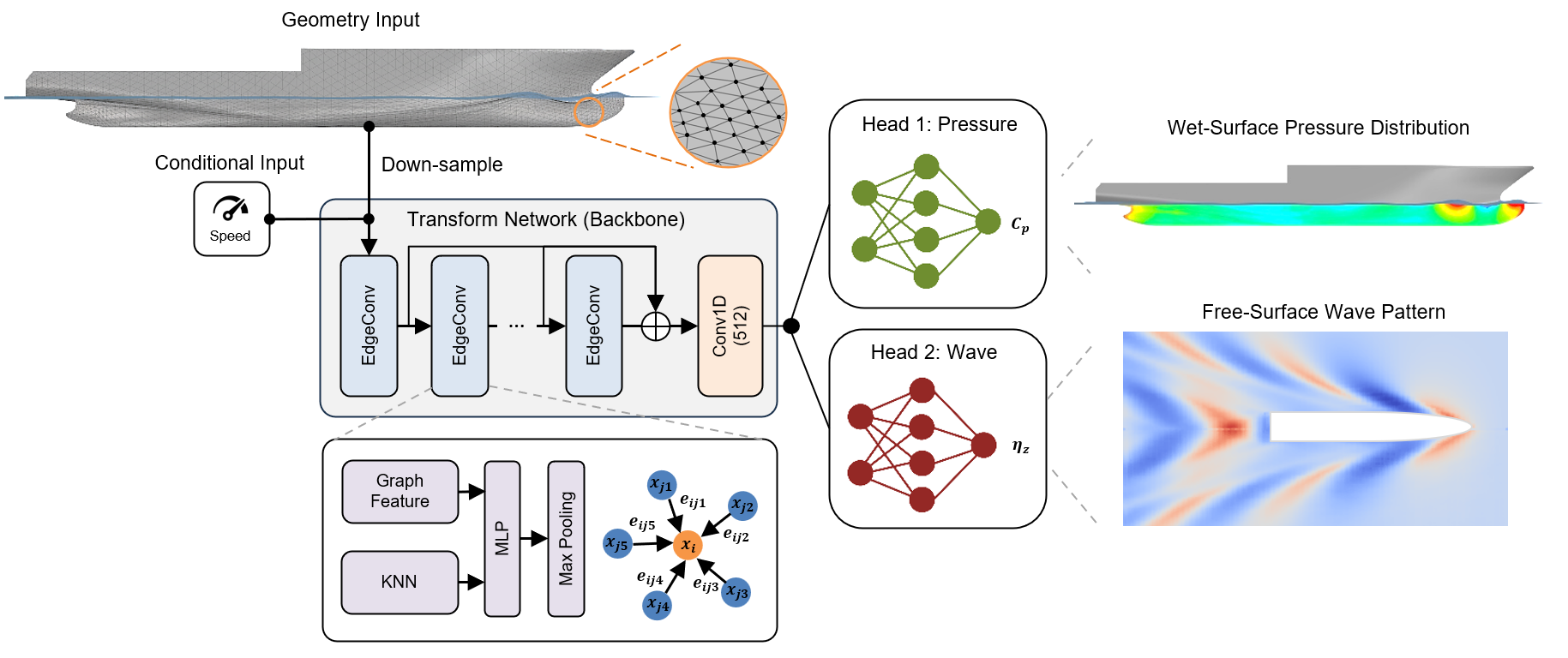}
\caption{The proposed multi-head architecture adapted from \cite{DrivAerNetPP}. An input point cloud is processed by a shared regDGCNN backbone to produce a latent feature representation. This representation is then decoded by two specialized heads to predict the hull surface pressure and the 2D free surface wave map.}
\label{fig:architecture_workflow}
\end{figure*}

\subsection{Data Preprocessing}
Each simulation case provides two data modalities: (i) a hull surface point cloud with associated pressure coefficient values, and (ii) a free-surface elevation field. To ensure consistency and comparability across cases, the following preprocessing protocol was applied:

\textbf{Geometric alignment.} All hull geometries were placed in a consistent global reference frame with the origin located at the midship, centerline, and calm-water plane: $X=L_{pp}/2, \;Y= waterline, \; Z= centerline.$ No additional geometric normalization (scaling by $\Lpp$) was applied because all variants belong to a common yacht class with fixed principal dimensions; thus coordinates remain in meters, and the operating condition is provided separately via $\Froude$.

\textbf{Hull point sampling.} The regDGCNN backbone operates on point-cloud representations and is therefore mesh agnostic, requiring only a sufficient number of surface samples rather than a fixed connectivity or discretization. From each watertight hull surface, $N_p = 1024$ points were sampled quasi-uniformly over the wetted area using a stratified random sampling strategy. This approach reduces bias introduced by non-uniform mesh tessellation and ensures that all regions of the hull contribute to the learned representation. While point-based methods are robust to changes in surface triangulation, adequate spatial coverage remains important to capture localized geometric features such as bow curvature and shoulder transitions. The influence of sampling density on predictive accuracy is examined explicitly in the ablation study in Section~4.1.

\textbf{Free-surface representation.} The free-surface wave elevation field was interpolated from the solver’s mesh grid onto a fixed Eulerian grid of resolution $(N_x,N_z)$. This yields a single-channel wave height map aligned with the ship-fixed frame for each case.

\textbf{Target normalization.} Both the hull pressure coefficient $(C_p)$ and the wave elevation $(\eta)$ were standardized to zero mean and unit variance using dataset-level statistics computed on the training set:
\footnotesize
\begin{equation*}
   C_p=(C_p-\mu_{cp})/\sigma_{cp}, \quad      \eta= (\eta-\mu_{\eta})/\sigma_{\eta}  
\end{equation*}
\normalsize
The model predicts in this normalized space, and losses are computed against standardized targets. For reporting and visualization, predictions are denormalized back to physical units. The dataset-level training statistics used for standardization are: $\mu_{cp}{=}0.53$, $\sigma_{cp}{=}0.5404$, $\mu_{\eta}{=}{-}5.94\!\times\!10^{-3}$\,m, $\sigma_{\eta}{=}3.949\!\times\!10^{-2}$\,m. This preprocessing ensures scale-invariant training losses, physically interpretable outputs at inference, and stability across all geometries in the dataset.


\begin{figure*}[t]
  \centering
  \begin{subfigure}[t]{0.24\textwidth}
    \centering
    \includegraphics[width=\linewidth]{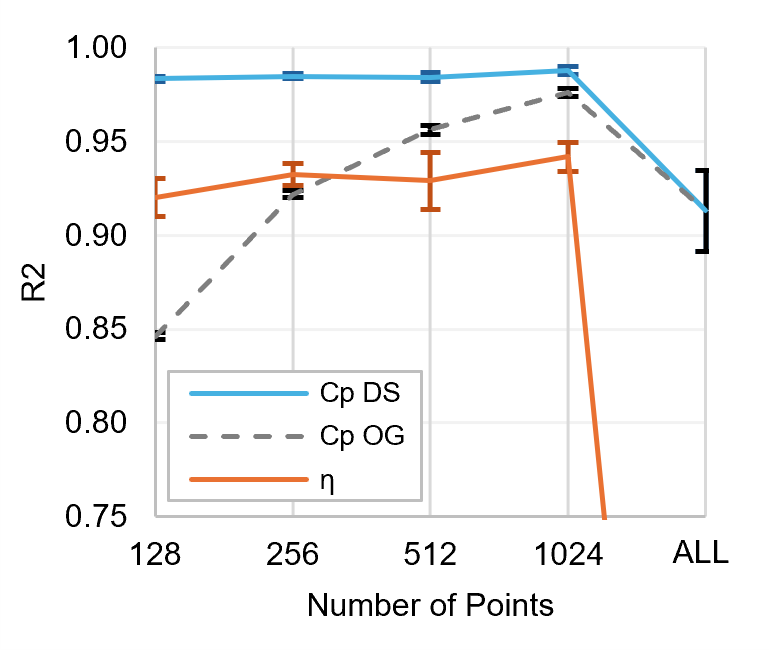}
    \caption{R\textsuperscript{2} vs points}
    \label{fig:grid-a}
  \end{subfigure}\hfill
  \begin{subfigure}[t]{0.24\textwidth}
    \centering
    \includegraphics[width=\linewidth]{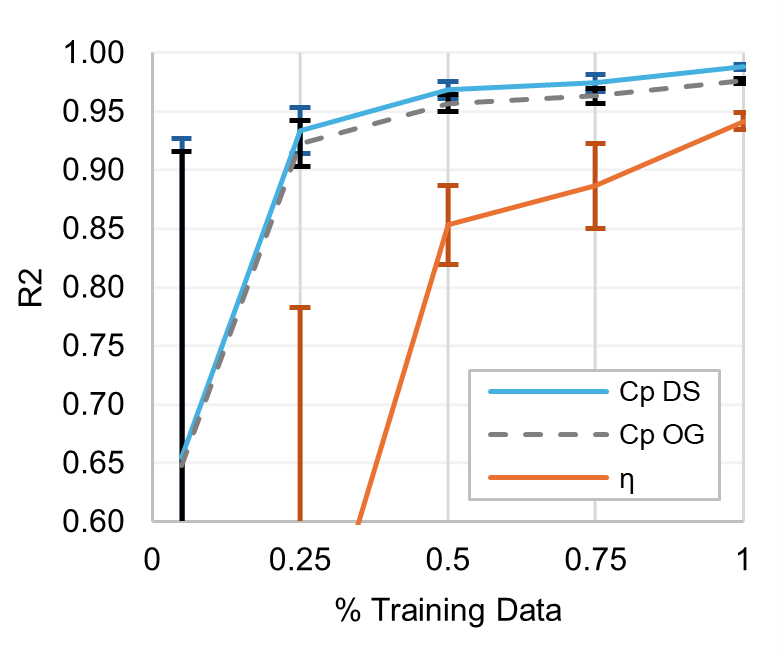}
    \caption{R\textsuperscript{2} vs \% data}
    \label{fig:grid-b}
  \end{subfigure}\hfill
  \begin{subfigure}[t]{0.24\textwidth}
    \centering
    \includegraphics[width=\linewidth]{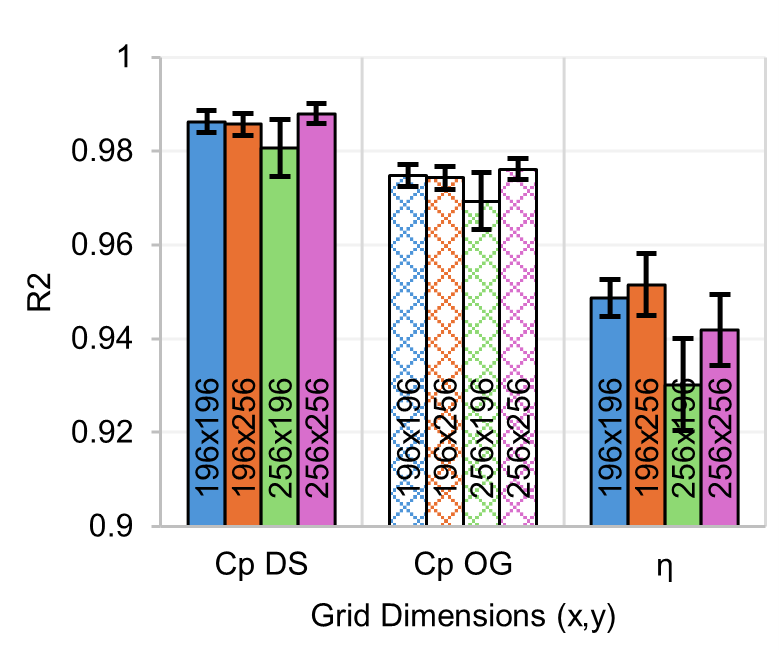}
    \caption{R\textsuperscript{2} vs grid}
    \label{fig:grid-c}
  \end{subfigure}\hfill
  \begin{subfigure}[t]{0.24\textwidth}
    \centering
    \includegraphics[width=\linewidth]{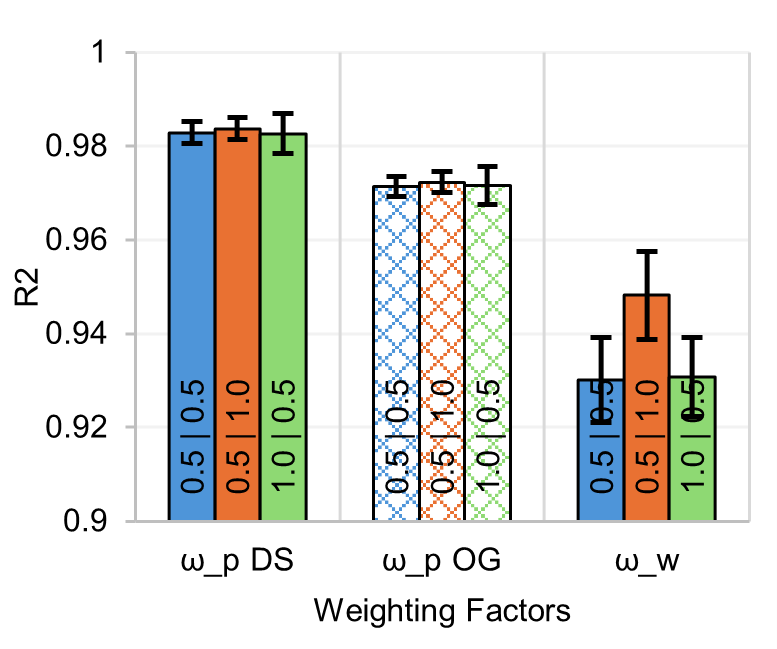}
    \caption{R\textsuperscript{2} vs weights}
    \label{fig:grid-d}
  \end{subfigure}

  \vspace{0.8em} 

  \begin{subfigure}[t]{0.24\textwidth}
    \centering
    \includegraphics[width=\linewidth]{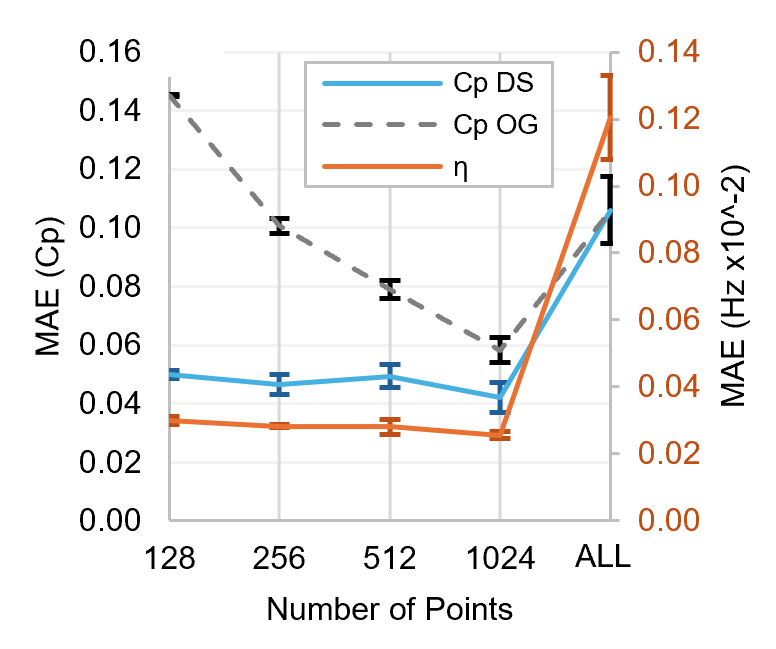}
    \caption{MAE vs points}
    \label{fig:grid-e}
  \end{subfigure}\hfill
  \begin{subfigure}[t]{0.24\textwidth}
    \centering
    \includegraphics[width=\linewidth]{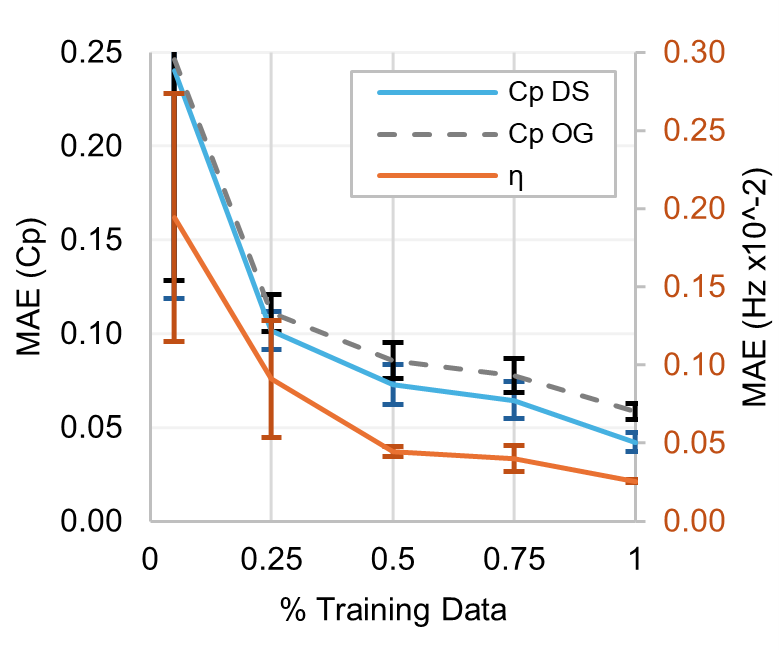}
    \caption{MAE vs \% data}
    \label{fig:grid-f}
  \end{subfigure}\hfill
  \begin{subfigure}[t]{0.24\textwidth}
    \centering
    \includegraphics[width=\linewidth]{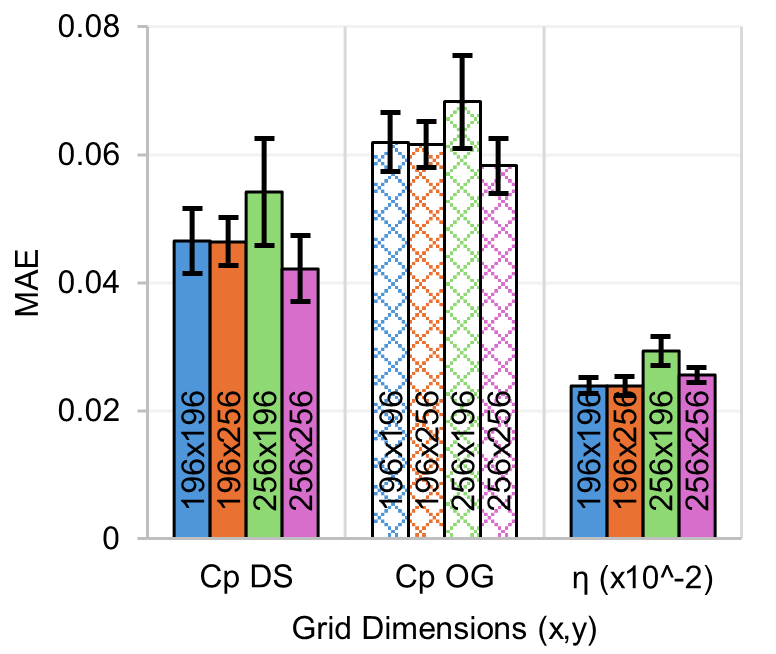}
    \caption{MAE vs grid}
    \label{fig:grid-g}
  \end{subfigure}\hfill
  \begin{subfigure}[t]{0.24\textwidth}
    \centering
    \includegraphics[width=\linewidth]{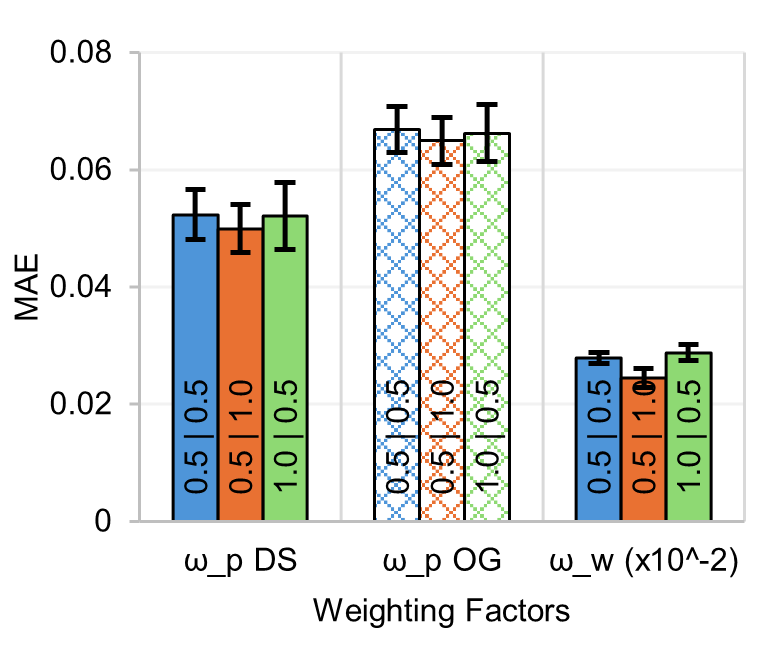}
    \caption{MAE vs weights}
    \label{fig:grid-h}
  \end{subfigure}
  \caption{Learning-test curve summaries across ablation studies.  (a–d) vary the number of points and training-data fraction; (e–h) vary grid resolution and loss-weight factors.  Error bars denote 95\% confidence intervals across 3 random seed variations. DS denotes Cp evaluated on down-sampled geometry points, while OG denotes Cp evaluated on the original full geometry, reconstructed via interpolation from the reduced point set, thereby capturing reconstruction error. The $\eta$ curve shows wave elevation prediction performance.}
  \label{fig:big-2x4}
\end{figure*}

\subsection{Multi-head regDGCNN Model}\label{sec:regDGCNN}

Our work adapts the regression-based Dynamic Graph Convolutional Neural Network (regDGCNN), an architecture that has demonstrated strong performance in predicting aerodynamic quantities directly from 3D point cloud representations of complex geometries \cite{Elrefaie_2025}. The regDGCNN model builds upon the Dynamic Graph CNN (DGCNN) framework proposed by Wang et al. \cite{WangDGCNN}, extending it from classification to continuous regression tasks. Its central building block is the EdgeConv operator, which is designed to capture local geometric structure directly from unstructured point sets.

Unlike grid-based convolutional neural networks, EdgeConv operates on a graph constructed from the point cloud. For each point, a local neighbourhood is defined by identifying its $k$ nearest neighbors in feature space. Edge features are formed from the relative differences between the central point and its neighbours and processed through a shared multilayer perceptron, followed by a symmetric aggregation operation such as max pooling \cite{WangDGCNN}. The neighbourhood graph is recomputed dynamically at each layer, allowing points to be grouped based on learned feature similarity rather than fixed spatial proximity. This dynamic construction enables the network to capture both fine-scale local geometry and increasingly global shape context.

The proposed multi-head architecture is illustrated in \Cref{fig:architecture_workflow}. A detailed architectural breakdown can be seen in Appendix \ref{app:architecture}. An input hull point cloud is processed by a shared regDGCNN backbone to produce a latent geometric representation, which is then decoded by two specialized heads for predicting hull-surface pressure and free-surface wave elevation.

\textbf{Shared feature extraction backbone.}  
The backbone ingests the sampled hull point cloud and processes it through three successive EdgeConv layers \cite{WangDGCNN}. Each layer dynamically constructs a $k$-nearest-neighbour graph and updates the point-wise feature representation through EdgeConv operations and non-linear activations. The resulting point-wise features are aggregated using a symmetric pooling operation to produce a fixed-size latent embedding that encodes the global hull geometry. The embedding dimension ($1024$) and neighbourhood size ($k = 40$) were selected to balance expressive capacity and computational cost, following prior regDGCNN studies and preliminary sensitivity tests.

\textbf{Operating condition input.}  
Each simulation case includes a scalar operating condition represented by the Froude number, $\Froude = U/\sqrt{g\Lpp}$. To condition the learned representation on speed, the Froude number is appended as an additional feature channel to every input point, such that each point is represented by $(x, y, z, \Froude)$. This formulation enables a single network to generalize across multiple operating conditions while maintaining a consistent geometric representation.

\textbf{Multi-head output formulation.}  
The latent embedding produced by the shared backbone is provided as input to two task-specific decoder heads. This multi-head formulation allows the network to exploit a common geometry-conditioned representation while learning specialized mappings for distinct physical outputs.

\textbf{Pressure prediction head.}  
The pressure head predicts the pressure coefficient $C_p$ at each sampled hull point. It combines the global latent embedding with the corresponding point-wise features from the final EdgeConv layer and processes the concatenated features through a sequence of one-dimensional convolutional layers. This design allows each pressure prediction to depend on both the local geometric context of the point and the global hull shape.

\textbf{Wave prediction head.}  
The wave head predicts the free-surface elevation field as a two-dimensional image in the ship-fixed reference frame. The global latent embedding is first passed through a fully connected layer and reshaped into a low-resolution feature map. This feature map is then progressively upsampled to the target resolution using a series of transposed convolutional blocks, following a decoder architecture analogous to U-Net \cite{UNet}. This design enables the network to map global geometric features to spatially distributed wave patterns while preserving the large-scale structure of the wake.

\renewcommand{\arraystretch}{1.25}
\begin{table}[t]
\centering
\captionsetup{width=0.95\linewidth}
\footnotesize
\resizebox{\columnwidth}{!}{%
\begin{tabular}{@{}p{0.33\linewidth} p{0.66\linewidth}@{}}
\toprule
\textbf{Parameter} & \textbf{Value} \\ \hline
Backbone & regDGCNN (embed=1024, k=40) \\
Heads & Pressure (3D), Wave (2D) \\
Model Params & $\sim$20M \\
Hull Vertices & 1220 – 4308 ($\mu$=2668) \\
Batch / Epochs & 4 / 100 \\
Optimizer & SGD; learn rate $10^{-3}$ \\
Scheduler & ReduceLROnPlateau(0.1, 10) \\
Dropout & 0.4 \\
Hardware & NVIDIA T4 (16--32\,GB VRAM) \\ \hline
\multicolumn{2}{l}{\textbf{Hyperparameter Sweep}}  \\ \hline
Number of points &	\{128, 256, 512, 1024\} \\
Data \% & \{0.05, 0.25, 0.50, 0.75, 1.0\} \\
Wave grid (x, y) &	\{256x256 , 196x256, 196x256, 196x196\}\\
Weights of Loss (w) &	\{0.5$\mid$0.5 , 1.0$\mid$0.5 , 1.0$\mid$0.5\}\\ \hline
\multicolumn{2}{l}{\textbf{Ablation}}  \\ \hline
Wave Loss & \{MSE, Hybrid\} \\
\bottomrule
\end{tabular}
}
\vspace{5pt}
\caption{Architecture and training configuration (primary setting).}
\label{tab:hparams}
\vspace{-15pt}
\end{table}

\begin{figure*}[t]
\centering
\includegraphics[width=\textwidth]{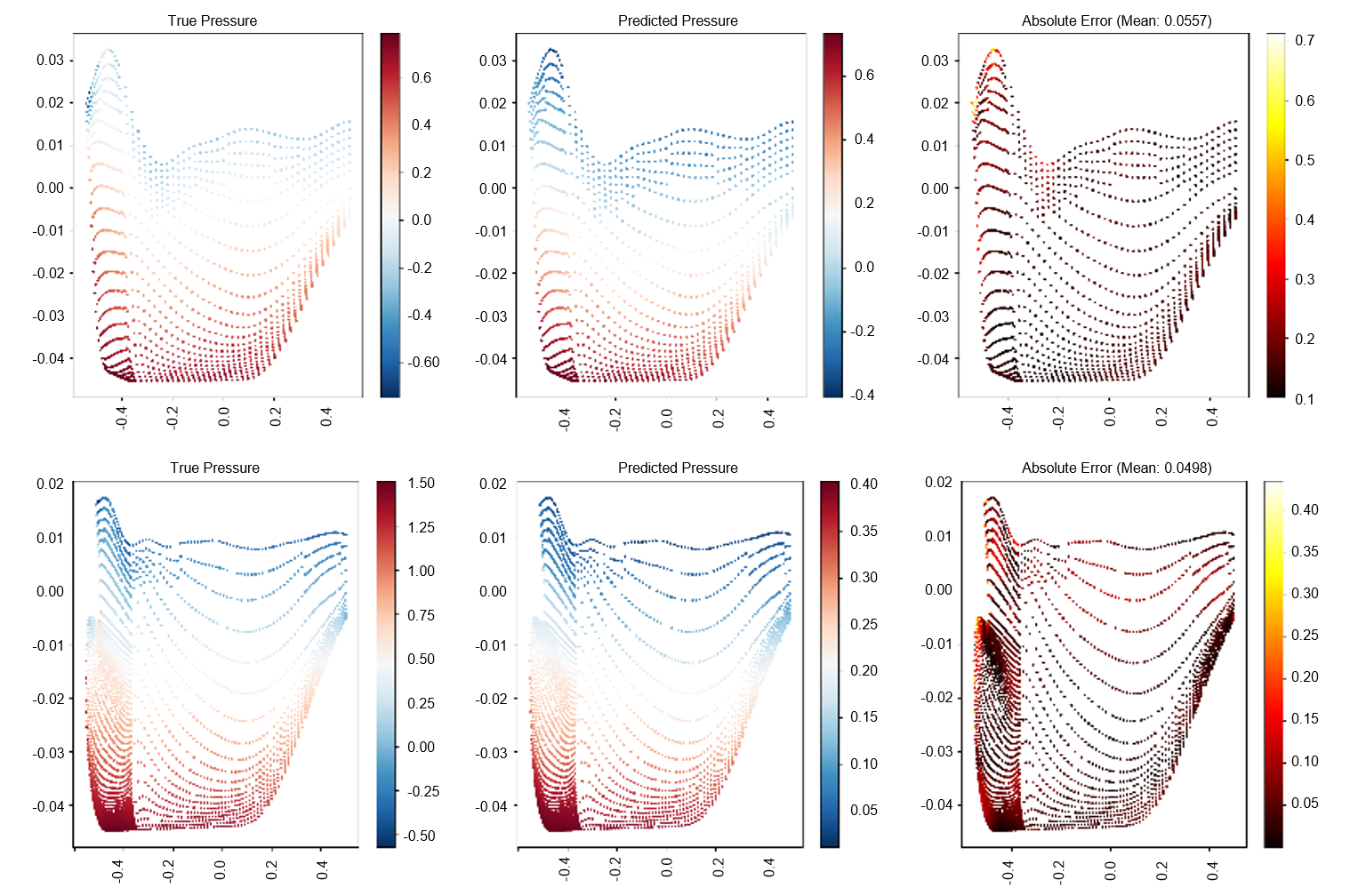}
\caption{non-dimensionalized pressure ($C_P$) projected on the 2D ZY-Plane: truth (left), prediction (middle), absolute error (right) for bulb (bottom) and straight (top) bow variants.}
\label{fig:pressure}
\end{figure*}

\begin{figure*}[t]
\centering
\includegraphics[width=\textwidth]{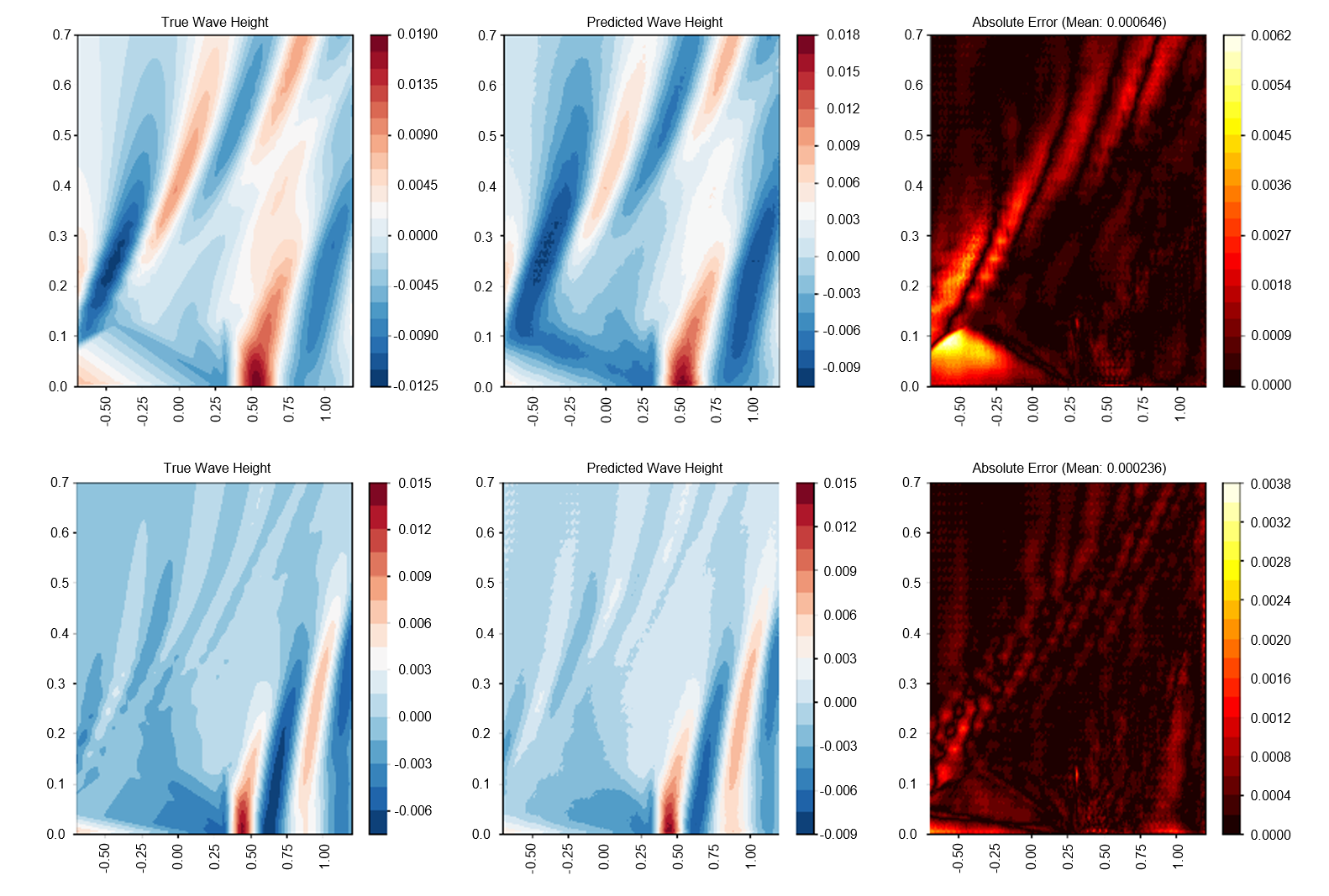}
\caption{Normalized 2D free-surface elevation ($\eta$): truth (left), prediction (middle), absolute error (right) for bulb (bottom) and straight (top) bow variants.}
\label{fig:waves}
\end{figure*}

\subsection{Training and evaluation}

The model is trained end-to-end using a composite loss that jointly optimizes the pressure and wave prediction tasks through a shared backbone. The total training objective is defined as
\begin{equation}
\mathcal{L}_{\text{total}} = \lambda_p \, \mathcal{L}_{\text{pressure}} + \lambda_w \, \mathcal{L}_{\text{wave}},
\end{equation}
where $\lambda_p$ and $\lambda_w$ are empirically selected weighting factors that balance the relative contributions of the two outputs. The pressure loss $\mathcal{L}_{\text{pressure}}$ is defined as the mean squared error between predicted and reference pressure coefficients at the sampled hull points.

\textbf{Wave loss formulation.}  
The free-surface wave field is optimized using a hybrid objective that combines complementary error measures:
{\footnotesize
\begin{equation}
\mathcal{L}_{\text{wave}} =
\alpha \lVert \hat{\eta} - \eta \rVert_1
+ \beta \bigl(1 - \mathrm{SSIM}(\hat{\eta}, \eta)\bigr)
+ \gamma \lVert \nabla \hat{\eta} - \nabla \eta \rVert_1,
\label{eq:wave_loss}
\end{equation}
}

where $\hat{\eta}$ denotes the predicted wave elevation map and $\eta$ the reference solution. The first term promotes pixel-wise accuracy while remaining less sensitive to large localized errors than an $L_2$ objective. The second term enforces structural similarity between predicted and reference wave patterns using the Structural Similarity Index Measure (SSIM) \cite{WangSSIM}, which compares local luminance, contrast, and structural information over sliding windows. The third term penalizes discrepancies in spatial gradients, computed using Sobel filters in the streamwise and transverse directions, and encourages accurate prediction of wave crests and troughs. 
The additive weighting coefficients ($\alpha, \beta, \gamma$) were set to equal values for all experiments in order to isolate architectural and representational effects. While task-specific tuning may further improve accuracy, such optimization is outside the scope of the present study.

\textbf{Optimization details.}  
Training is performed using stochastic gradient descent with learning-rate scheduling and regularization as summarized in \Cref{tab:hparams}. Hyperparameter values are informed by the DrivAerNet++ baseline, reflecting the similarity of the underlying geometry-conditioned regression task. The multi-head formulation requires careful loss balancing to prevent domination of one task over the other, as discussed in the ablation study.

\textbf{Evaluation metrics.}  
Model performance is assessed on an unseen test set using standard regression metrics. For both hull pressure and free-surface wave predictions, we report the mean absolute error (MAE), mean squared error (MSE), and the coefficient of determination ($R^2$). Metrics are computed on a per-sample basis and summarized by their mean, standard deviation, minimum, and maximum values across the test dataset.

\section{Results}\label{results}
\subsection{Hyperparameter Optimization}
A targeted sensitivity study was conducted to evaluate the effects of point sampling density, training data volume, wave-grid resolution, and loss weighting. \Cref{fig:big-2x4} summarizes these ablations with mean performance and 95\% confidence intervals over the test set. Three response types are considered: pressure prediction accuracy on the downsampled point cloud used during training, pressure accuracy after re-interpolation to the full-resolution hull surface, and free-surface wave elevation accuracy. Since the regDGCNN backbone operates on a fixed number of sampled points, training is necessarily performed on a reduced representation of the hull. Adequate sampling density is therefore essential to retain sufficient geometric information for accurate reconstruction on the full domain.

A consistent performance gap is observed between pressure accuracy on the sampled point set and on the reconstructed full-resolution geometry. This gap decreases monotonically with increasing point count and becomes negligible at approximately $N_p = 1024$, indicating that this sampling density preserves the relevant geometric features. Both $R^2$ and MAE follow the same trend, with performance degrading as point count decreases and reconstruction error increases. With respect to training data volume, performance saturates at roughly 50\% of the available dataset, corresponding to approximately 80 hull-speed samples, demonstrating strong data efficiency. Training on the full dataset using a single batch leads to degraded performance, which is attributed to optimization limitations rather than model capacity.

Wave-grid resolution exhibits comparatively weak sensitivity within the tested range. However, changes to wave-related hyperparameters and loss weighting affect both wave and pressure accuracy, reflecting the shared latent representation in the multi-head architecture. In particular, loss weighting plays a critical role, as imbalanced task objectives can bias the learned representation toward one output at the expense of the other. These results emphasize the importance of careful hyperparameter selection in jointly trained geometric surrogate models.

\raggedbottom

\subsection{Quantitative Accuracy}\label{sec:quantitative}
Using the out comes of the hyper-parameter investigation a final model was trained and evaluated. The loss curves of the training process can be seen in Appendix \ref{app:loss}.  On the test set (63 cases), we observe high accuracy for both outputs (\cref{tab:metrics}). Pressure achieves $R^2\!=\!0.981\!\pm\!0.014$ (min 0.908, max 0.994). Waves achieve $R^2\!=\!0.912\!\pm\!0.070$ (min 0.656, max 0.985). Mean absolute and squared errors are also reported.

\begin{table*}[t]
\centering
\begin{tabular}{@{}l S S S S@{}}
\toprule
Metric & {Mean \((\mu)\)} & {Std \((\sigma)\)} & {Min} & {Max} \\
\midrule
MAE$_\mathrm{pressure}$ (-) & 5.00e-02 & 1.40e-02 & 2.69e-02 & 8.80e-02 \\
$R^2_\mathrm{pressure}$ (\%) & \text{98.1} & 1.44 & 90.8 & 99.4 \\ \midrule
MAE$_\mathrm{wave}$ & 3.69e-04 & 1.45e-04 & 1.64e-04 & 7.50e-04 \\
$R^2_\mathrm{wave}$ (\%) & 91.2 & 7.04 & 65.6 & 98.5 \\
\bottomrule
\end{tabular}
\caption{Test-set metrics (mean $\mu$, std $\sigma$, min, max).}
\label{tab:metrics}
\end{table*}

While aggregate metrics confirm what the model achieves, a qualitative analysis is necessary to understand how and where it performs well.

\textbf{Hull pressure distribution.} \Cref{fig:pressure} visualizes the model's predictions on two representative but challenging cases from the test set: a straight bow hull at high speed and a bulbous bow hull. The model correctly identifies the primary features of the pressure field: the high-pressure stagnation point at the bow, the low-pressure region at the shoulder leading to pressure recovery, and the overall distribution along the waterline.

The error maps are particularly revealing. They show that the largest errors are not random but are systematically located in regions of maximum geometric complexity and pressure gradients—specifically, the sharp curvature at the bulb's leading edge and its intersection with the hull. This is physically intuitive; these are locations where the flow changes most rapidly. The fact that the model's errors are localized to these challenging areas while the bulk of the hull surface is predicted with very high accuracy demonstrates its ability to learn the fundamental pressure-geometry relationship.

\textbf{Free surface wave pattern.} The wave predictions in \Cref{fig:waves} showcase the model's ability to capture complex, spatially distributed phenomena. The model successfully reproduces the characteristic Kelvin wake structure, including the divergent and transverse wave systems. More impressively, it learns to differentiate the subtle but important changes in the wave pattern caused by geometric modifications, such as the wave-cancelling effect of a well-designed bulbous bow.

The error maps highlight that the largest discrepancies occur within the first wave trough immediately aft of the bow. This region is hydrodynamically complex, often on the verge of wave breaking, and is highly sensitive to the exact stagnation pressure at the bow. The model's slight under-prediction of the bow stagnation pressure likely translates directly to this slight under-prediction of the initial wave trough amplitude.

\begin{figure}[t]
\centering
    \begin{subfigure}[b]{0.95\columnwidth}
        \centering
        \includegraphics[width=0.95\linewidth]{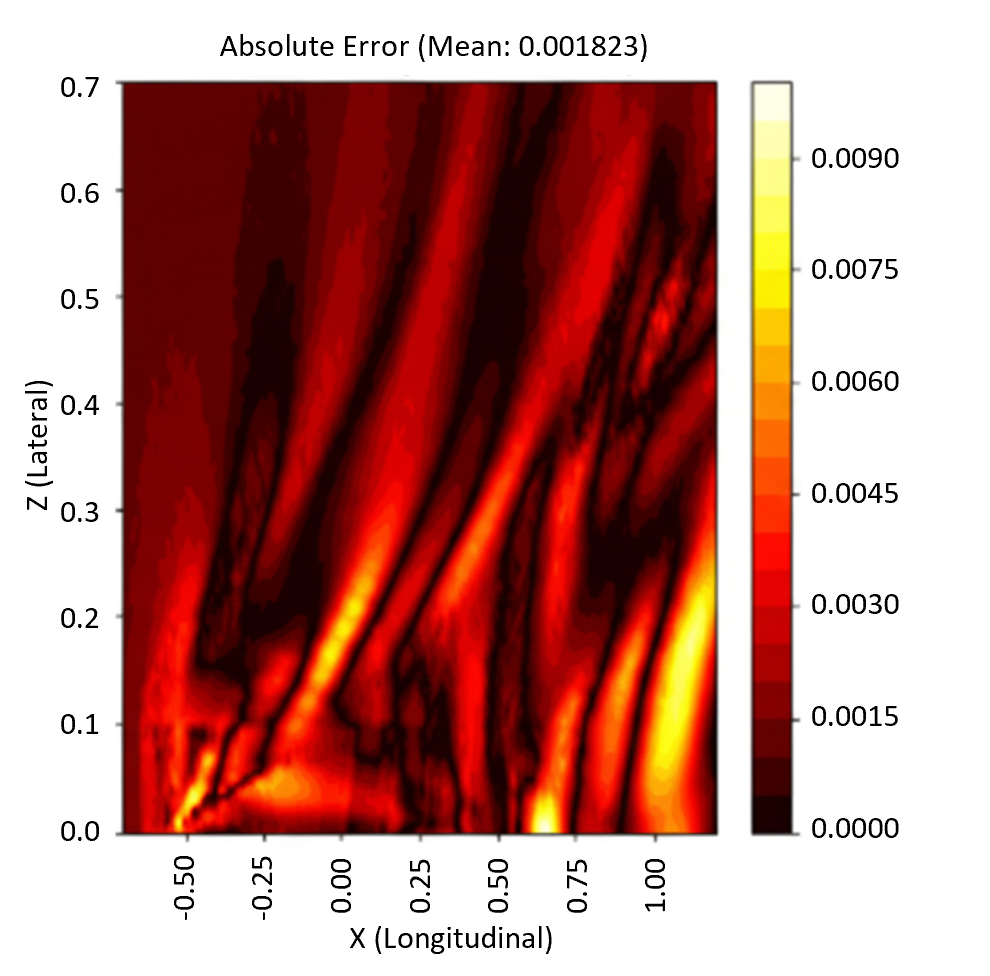}
        \caption{\label{fig:ablation_mse}Wave prediction trained with MSE.}
    \end{subfigure}
    \quad
    \begin{subfigure}[b]{0.95\columnwidth}
        \centering
        \includegraphics[width=0.95\linewidth]{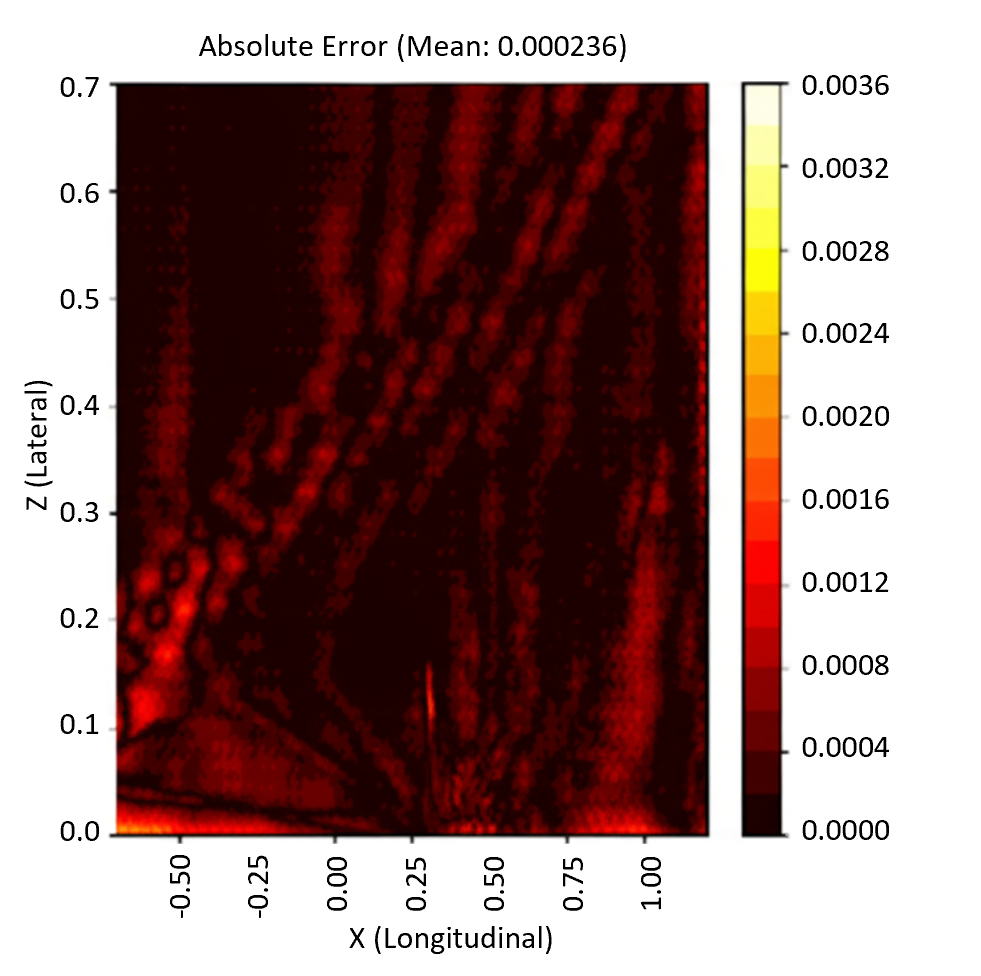}
        \caption{\label{fig:ablation_comp}  Wave prediction trained with hybrid loss.}
    \end{subfigure}

    \caption{Qualitative loss ablation for the wave head: standard MSE leads to a trivial solution, while the hybrid loss improves structural fidelity.}
    \label{fig:ablation_loss}
    \vspace{-5pt}
\end{figure}

\subsection{Ablations and loss design}\label{sec:ablation}
The choice of loss function was found to be the single most critical factor in achieving accurate wave predictions. An ablation study training the model with a standard MSE loss for the wave head resulted in complete training failure, yielding a mean R² of -0.360 (\Cref{tab:lossablation}). A negative R² signifies a model that performs worse than simply predicting the mean of the data.

\renewcommand{\arraystretch}{1.25}
\begin{table}[t]
\centering
\captionsetup{width=0.95\linewidth}
\footnotesize
\resizebox{\columnwidth}{!}{%
\begin{tabular}{@{}l S[table-format=+1.3] S[scientific-notation=true,table-format=1.2e-2]@{}}
\toprule
Wave Loss & {$R^2_{wave}$} & {$\mathrm{MAE_{wave}}$} \\
\midrule
Single: MSE & -0.360 & 2.06e-03 \\
Hybrid: L1 + SSIM + Sobel & 0.912 & 3.70e-04 \\
\bottomrule
\end{tabular}
}
\vspace{5pt}
\caption{Loss ablation for wave prediction.}
\label{tab:lossablation}
\vspace{-10pt}
\end{table}

\Cref{fig:ablation_loss} shows the stark difference in performance. This failure is a direct consequence of the data's sparsity. MSE loss, by squaring the error, is highly sensitive to the large errors that inevitably occur when predicting the location of sharp, high-amplitude wave crests in a field that is otherwise near-zero. This forces the optimizer into a trivial local minimum, resulting in a blurry, over-smoothed output that minimizes the large-error penalty at the cost of all physical realism. In contrast, our composite loss (L1+SSIM+Sobel) is purpose-built for such tasks. The L1 term is less sensitive to large errors, the SSIM term enforces structural and perceptual similarity, and the Sobel term explicitly rewards the correct prediction of high-frequency features (steep wave gradients). This synergistic combination proved essential for guiding the model to a physically meaningful solution.

\subsection{Computational cost}
The primary engineering value of our model lies in its computational efficiency. A RAPID simulation requires approximately 3.8 minutes (228 s) on a single CPU core. In contrast, our trained model performs inference for a single case in 0.15s/iter on an NVIDIA T4 GPU. This represents a speedup of approximately 1500x, transforming the analysis from a batch-process workflow into a near-instantaneous, interactive tool. This acceleration fundamentally changes the paradigm of early-stage design, allowing for comprehensive parametric sweeps and real-time analyses that are impossible with traditional simulation tools. See \Cref{tab:speedup} for detailed summary.

\subsection{3D output representation and visualization}

Although the quantitative results in Sections \ref{results} are visualized using two-dimensional plots for clarity, ShipNet predicts fully three-dimensional physical quantities. The apparent dimensional reduction reflects a visualization choice and not a limitation of the underlying model or data pipeline.

The hull pressure prediction is defined on the wetted hull surface as a scalar field on a 3D manifold. Predictions are generated at arbitrary surface point locations and can be re-interpolated to the full-resolution hull geometry. The resulting pressure fields are exported as VTK-compatible datasets, allowing direct use in standard post-processing and visualization tools such as ParaView. The free-surface prediction represents a three-dimensional free surface in height-field form, $\eta(x,y)$, over a ship-fixed spatial domain. While shown here as 2D colormaps, the predicted field can be reconstructed as a full 3D surface mesh and similarly exported in VTK format for integration into existing CFD and hydrodynamic analysis workflows. Appendix \ref{app:3dviz} presents representative three-dimensional visualizations of both outputs, illustrating the full spatial structure captured by the model.

\renewcommand{\arraystretch}{1.25}
\begin{table}[t]
\centering
\captionsetup{width=0.95\linewidth}
\footnotesize
\resizebox{\columnwidth}{!}{%
\begin{tabular}{@{}l l l@{}}
\toprule
Comparison & Speed-up & Note \\
\midrule
GPU vs.\ CPU (train iter) & 2.7$\times$ & 0.15\,s vs.\ 0.40\,s/iter \\
RAPID vs.\ CPU inference & 550$\times$ & 228\,s vs.\ 0.40\,s \\
RAPID vs.\ GPU inference & 1500$\times$ & 228\,s vs.\ 0.15\,s \\
\bottomrule
\end{tabular}
}
\vspace{5pt}
\caption{Computational performance summary.}
\label{tab:speedup}
\vspace{-10pt}
\end{table}

\section{Limitations and Future Work}
Despite the promising results, we acknowledge several limitations that provide clear avenues for future research.
\begin{enumerate}[topsep=0pt, itemsep=0.0pt, leftmargin=*]
    \item \textbf{Data fidelity:} The model is trained exclusively on potential flow data. As a result, it learns to predict an inviscid flow field. The next logical step is to retrain and validate this architecture on a high-fidelity RANS-CFD dataset to demonstrate its capability to capture viscous effects, such as boundary layer development and transom wake dynamics. 
    \item \textbf{Generality of design space:} The training data was limited to a specific family of slender yacht hulls. While the model showed strong generalization within this class, its performance on hydrodynamically different vessels (full-form tankers) is not guaranteed. Future work should involve expanding the dataset to include a more diverse range of vessel types. The complex flow around transom sterns, which was not varied in this study, represents a particularly important area for future investigation.
    \item \textbf{Geometric complexity:} The current study was limited to bare-hull geometries. The influence of appendages such as rudders, keels, and stabilizers on the pressure field and wave system is significant and should be incorporated in future datasets.
    \item \textbf{Architectural and hyperparameter optimization:} The proposed architecture, while effective, is complex. As an adaptation of an existing model, it has not been optimized from the ground up for this specific task. A systematic study of architectural choices (number of layers, neurons) and a more extensive hyperparameter search could yield further improvements in accuracy and efficiency. Additionally, investigating the sensitivity of the model to the number of input points sampled from the hull is a key area for future analysis.
    \item \textbf{Physical constraints:} Fundamentally, the model does not explicitly enforce physical constraints such as mass conservation or turbulence-model consistency, which may lead to non-physical artifacts, particularly in regions with sparse training data. It would be interesting to perform extrapolation studies and ablate physics informed loss to check whether fundamental laws are retained or not.
\end{enumerate}

\section{Conclusion}
In this study, we addressed the critical need for rapid and accurate hydrodynamic analysis in the early stages of ship design. We have successfully developed and validated a novel multi-head geometric deep learning architecture capable of predicting both the hull surface pressure distribution and the free surface wave pattern directly from a 3D point cloud representation of the vessel. The core of our approach is a graph neural network backbone that learns a rich, physically-informed latent representation of the hull geometry, which is then decoded by two specialized heads for each prediction task.

The proposed model demonstrated exceptional performance on an unseen test set, achieving a mean R² of 0.981 for pressure and 0.912 for wave elevation. This high fidelity is complemented by a transformative increase in computational speed, delivering a speedup of over 1500x compared to the conventional potential flow solver used for data generation. This work serves as a successful proof-of-concept, demonstrating that graph-based neural networks are a powerful and effective tool for creating surrogate models for complex fluid-structure interaction problems.
\subsection{Broader impact and implications}
The primary impact of this research is its potential to reshape the naval architectural design spiral. By reducing the simulation latency from minutes or hours to milliseconds, our approach enables a truly interactive and data-rich design process. This allows for the systematic exploration of vast design spaces, which is infeasible with traditional simulation workflows.

For the maritime industry, the widespread adoption of such validated tools could yield significant long-term benefits. Enabling the design of more hydrodynamically efficient fleets would directly contribute to substantial fuel savings and the reduction of greenhouse gas emissions. Furthermore, the ability to generate multi-physics insights from geometry in near real-time is a foundational technology for the creation of vessel "digital twins," which can be leveraged for both advanced design and on-board operational support systems. The methodology itself is also broadly applicable, providing a framework for tackling similar challenges in the aerospace, renewable energy, and biomedical engineering sectors

\section*{Acknowledgments}
This research was made possible through funding from the Cooperative Research Ships (CRS) fund. The authors wish to express their gratitude for this essential support.

\raggedbottom

\bibliographystyle{unsrtnat}
\bibliography{references}

\clearpage            
\onecolumn            
\appendix

\section{Detailed ShipNet Architectural Breakdown}
\label{app:architecture}

Appendix \ref{app:architecture} provides a detailed schematic of the ShipNet architecture, expanding on the overview presented in Section \ref{sec:regDGCNN}. The figure illustrates the full data flow from the input hull point cloud and operating condition through the shared regDGCNN backbone and into the two task-specific decoder heads.

The diagram highlights how local point-wise features and global geometric embeddings are combined for hull pressure prediction, as well as how the global latent representation is reshaped and decoded into a spatially distributed free-surface elevation field. Layer dimensions, pooling operations, and loss associations are included to clarify implementation details that are omitted from the main text for brevity.

\begin{figure}[H]   
  \centering
  \includegraphics[width=\linewidth]{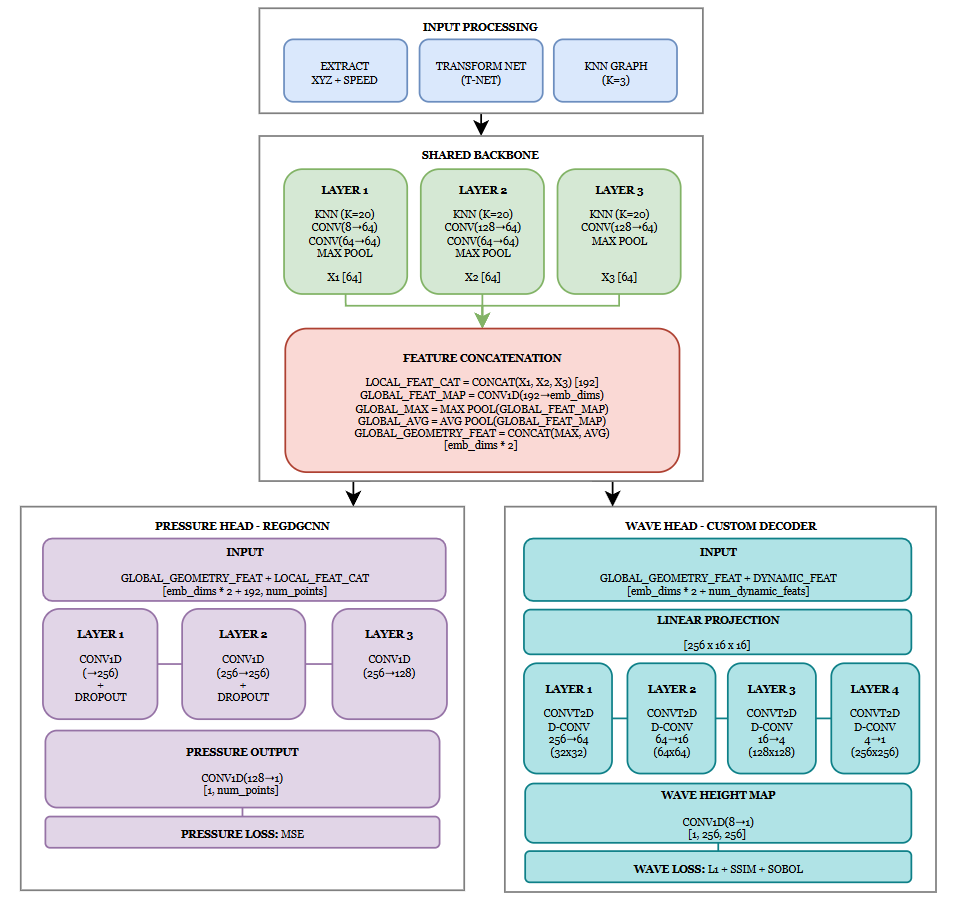}
  \caption{Detailed Architectural Breakdown.}
  \label{fig:architecture}
\end{figure}

\clearpage

\section{Training and Validation Loss Curves}
\label{app:loss}

Appendix \ref{app:loss} presents the training and validation loss histories for the final selected model, including the total composite loss as well as the individual pressure and wave loss components. These curves demonstrate stable convergence behaviour and indicate that neither prediction task dominates the shared optimization process.

The loss contribution ratios further illustrate the effectiveness of the chosen loss weighting strategy, supporting the ablation results discussed in Section \ref{sec:ablation} and confirming balanced multi-task learning throughout training.

\begin{figure}[H]
  \centering
  \includegraphics[width=\linewidth]{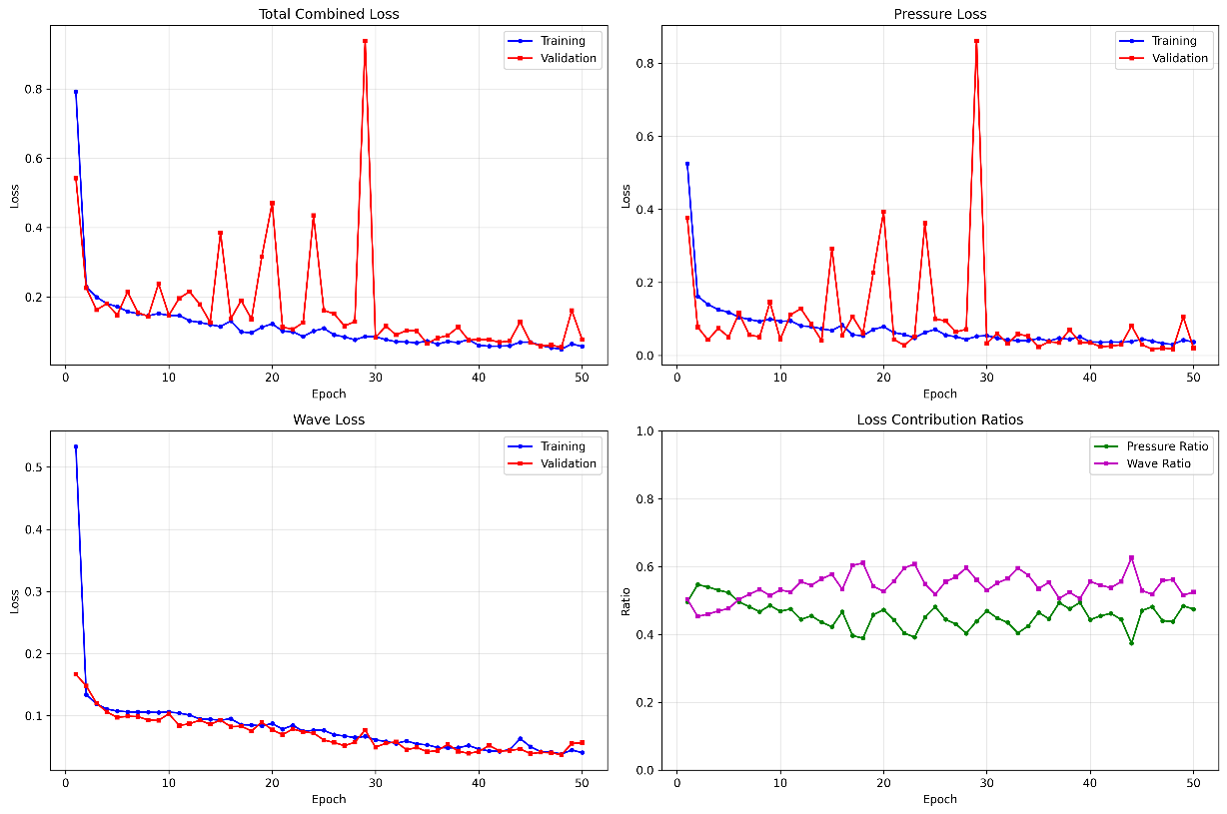}
  \caption{Training and Validation Loss Curves.}
  \label{fig:loss_curves}
\end{figure}

\clearpage

\section{Three-Dimensional Visualization of Predicted Fields}
\label{app:3dviz}

Appendix \ref{app:3dviz} provides representative three-dimensional visualizations of the predicted hull pressure and free-surface fields, reconstructed from the model outputs and exported in VTK format. These renderings complement the two-dimensional visualizations shown in the main text and demonstrate that ShipNet produces fully three-dimensional, spatially resolved predictions.

Hull pressure is shown as a scalar field mapped onto the complete wetted hull surface, enabling inspection of localized features such as bow stagnation regions and pressure recovery along the shoulder. The free surface is reconstructed as a three-dimensional surface mesh from the predicted elevation field, illustrating the spatial structure of the wake and Kelvin wave system.

All visualizations were generated using standard post-processing tools (ParaView), highlighting the compatibility of the proposed pipeline with existing CFD visualization and analysis workflows.

\begin{figure}[H]
  \centering
  \includegraphics[width=\linewidth]{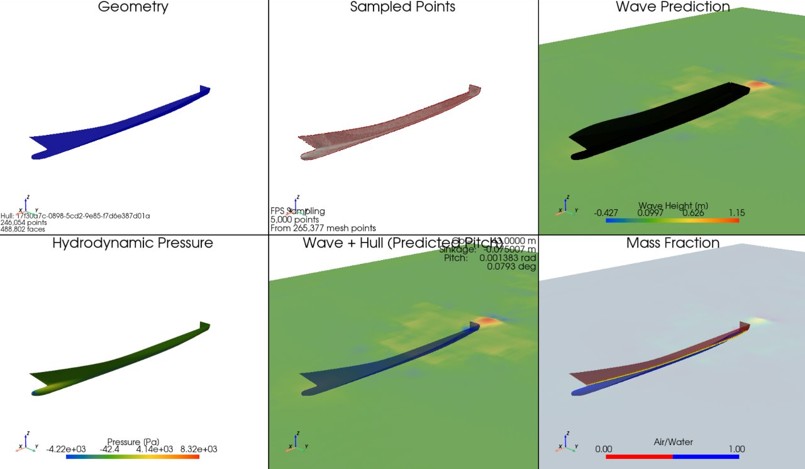}
  \caption{3D VTK representation visualized in ParaView.}
  \label{fig:3dvtk}
\end{figure}

\end{document}